\documentclass[12pt]{amsart}
\usepackage{amssymb}

\begin{document}
\title{Quantum computation with scattering matrices}
\author{ G. Giorgadze }
\address{Joint Institute for Nuclear Research, Dubna, Russia}
\email{ggk@jinr.ru}
\author{R. Tevzadze}
\address{Institute of Cybernetics,
         Georgian Academy of Sciences,
         Tbilisi, Georgia}%
\email{tevza@cybernet.ge}%

\thanks{To appear in ``Contemporary mathematics and its
applications''}

\maketitle

\bibliographystyle{plain}
\begin{abstract}
We discuss possible applications of the 1-D direct and inverse scattering
problem to design of universal quantum gates for quantum computation. The
potentials generating some universal gates are described.
\end{abstract}

\section{Introduction}

In this article we propose a theory of quantum scattering and notion of
unitary scattering matrix to formulate quantum input-output relations. This
differs from standard approach to this subject in which the quantum gates for
quantum computing are considered to be unitary time evolution operator for a
given, fixed time and with the Hamiltonian which describes the dynamics.

Representation of quantum gates as scattering matrices (S-matrices) may have
a physical realization. In present paper we study one-dimensional scattering
problem in electro-magnetic field. Varying electro-magnetic field and
momentum of the electron we get a 1-parametric family of S-matrices. We prove
that adjusting the electro-magnetic field we can deliver relevant 2-order and
4-order unitary matrix. It is well known that such matrices can be used as
universal gates for quantum computing \cite{bry1}.

Another type of a scattering problem emerges in two-level quantum systems
controlled by electric pulse. By a time-dependent Hamiltonian we also achieve
the representation of desired gates as scattering matrices. In both
approaches we use well-known inverse problem solutions to construct quantum
gates.

Application of the quantum scattering process to quantum computing was
studied in the works \cite{kk}, \cite{Lee}, \cite{mas}. Articles
\cite{giorgadze1}, \cite{giorgadze2} are devoted to the monodromic approach
to the quantum computing. In these works we construct the set of universal
gates by monodromy matrices of Fuchsian connections on holomorphic vector
bundles, such approach corresponds to holonomies of connections in the
holonomic quantum computing \cite{zanardi}, but holonomies do not depend  on
the path of integration. Unlike the monodromy matrices the scattering
matrices are encountered in physical experiments, moreover there exists a
one-to-one correspondence between the monodromy and scattering matrices
\cite{arnold} in one-dimensional case. In this article we describe potentials
and corresponding S-matrices which generate the set of universal gates.

The paper is organized as follows. In section 2 we discuss possible
applications of the 1-D electron scattering in an electro-magnetic potential
to the problem of design of universal quantum gates for quantum computation.
Using methodology of the inverse scattering problem we construct genuine
potentials to obtain some kind of universal gates as a scattering matrix. In
section 3 we consider the electric field interacting with two-level (or
four-level) quantum systems. Transitions from initial state to final state
are used as gates. The well known inverse scattering problem procedure allows
to obtain potentials that maintain some universal gates. Finally we consider
some potentials which allow to compute S-matrix as monodromy of the
corresponding Fuchsian system.

\section{Scattering on the line and universal gates}

Let us consider the stationary Schr\"odinger equation on the line
\begin{equation}\label{sch}
\frac{d^2}{dx^2}\psi(x)+(k^2+Q(x))\psi(x)=0, \text{  } x\in (
-\infty,\infty)
\end{equation}
where $Q(x)$ is a continuous potential function vanishing at infinity, i.~e.
$Q(x)\to 0,$ as $|x|\to\infty$.

Let $\varphi(x)$ be a solution of (\ref{sch}) which coincides with
$e^{-ikx}$ for $x\to-\infty.$ Its complex conjugate function
$\overline{\varphi}(x)$ also satisfies the equation (\ref{sch})
which coincides with $e^{ikx}$ as $x\to-\infty.$ Moreover we
denote by $\psi(x)$ and $\overline{\psi}(x)$ the solution of
(\ref{sch}) which coincides with $e^{ikx}$ and $e^{-ikx}$
respectively as $x\to\infty.$  It is clear that there exists
$2\times 2$ matrix $M(k)$ such that
\begin{equation}
(\varphi(x),\bar{\varphi}(x))=(\bar\psi(x),{\psi}(x))M(k).
\end{equation}

It is known that $M(k)\in SU(1,1)$ and this matrix is called monodromy matrix
\cite{arnold}. The matrix $M(k)$ can be represented as
$$
M(k)=\left(
\begin{array}{cccc}
a(k) &\bar{b}(k) \\
b(k) & \bar a(k)
\end{array}
\right),|a(k)|^2-|b(k)|^2=1.
$$
as an element of $SU(1,1)$. Therefore we have
$$
\begin{cases}
\varphi=a\bar{\psi}+b\psi \\
\bar{\varphi}=\bar{b}\bar\psi+\bar{a}\psi
\end{cases}
\Rightarrow
\begin{cases}
\bar\psi=\frac{1}{a}\varphi-\frac{b}{a}\psi \\
\bar{\varphi}=\bar{b}(\frac{1}{a}\varphi-\frac{b}{a}\psi)+\bar{a}\psi
=\frac{\bar{b}}{a}\varphi+\frac{1}{a}\psi
\end{cases}.
$$
Thus $(\bar{\psi},\bar{\varphi})=(\varphi,\psi)\widetilde{S}(k)$, where  $
\widetilde{S}(k)=\left (
\begin{array}{cccc}
\frac{1}{a} &\frac{\overline{b}}{a} \\
-\frac{{b}}{a} & \frac{1}{a}
\end{array}
\right ).
 $
Similarly $(\bar{\varphi},\bar{\psi})=(\varphi,\psi)S(k),$ where $
S(k)=\left (
\begin{array}{cccc}
\frac{\overline{b}}{a} &\frac{1}{a} \\
\frac{1}{a} &-\frac{{b}}{a}
\end{array}
\right ).
 $
The matrices $\widetilde{S}(k),S(k)$ are called \cite{newell}, \cite{fadeev}
the scattering matrices in various context.

We will use the notations $T(k)=\frac{1}{a(k)},$ $R(k)=\frac{b(k)}{a(k)}$
having an interpretation of transmission and reflection amplitudes
respectively. The particle comes by left-to-right with impulse $k>0$ goes
through the barrier with transmission probability $|T(k)|^2$ and goes back
with reflection probability $|R(k)|^2$, then output will be
$e^{ikx}+be^{-ikx}$ on left from the barrier, and $ae^{ikx}$ on the right
from the barrier.

We consider here only the case $S(k)$. Such matrix is symmetric and unitary
simultaneously (but it does not belong to $SU(2)$ in general) and many
interesting gates can be represented in such form. Let us consider the
mapping
$$
\tau:SU(1,1)\rightarrow U(2),\;\;
 \left (
\begin{array}{cccc}
a &\bar{b} \\
{b} & \bar a
\end{array}
\right )\mapsto\left (
\begin{array}{cccc}
\frac{\overline{b}}{a} &\frac{1}{a} \\
\frac{1}{a} &-\frac{{b}}{a}
\end{array}
\right ).
$$
The image of this mapping $\tau(SU(1,1))$ is the subset of the matrix in
$U(2)$, which can be represented as scattering matrix.

{\bf Remark 1.} Let initial states be denoted by $|0\rangle=e^{ikx}$,
$|1\rangle=e^{-ikx}$. If the scattering matrix has the form
$H=\frac{1}{\sqrt{2}} \left(
\begin{array}{cccc}
1&1 \\
1&-1\\
\end{array}
\right )$ then we obtain the states:
$(|0\rangle+|1\rangle)\sqrt{2}$ and
$(|0\rangle-|1\rangle)\sqrt{2}.$

Suppose that to write in quantum registers a natural number from 1 to $x$ an
 $n$-qubit is needed. We take $n$-th tensor product of Hadamard gates and obtain:
$$
(H\otimes...\otimes H)|0\rangle\otimes...\otimes |0\rangle
=\frac{1}{\sqrt{2^n}}H|0\rangle\otimes H|0\rangle\otimes...\otimes
H|0\rangle=
$$
$$
=\frac{1}{\sqrt{2^n}}(|0\rangle+|1\rangle)\otimes...\otimes
(|0\rangle+|1\rangle)=\frac{1}{\sqrt{2^n}}\sum_{x=0}^{2^n-1}|x\rangle.
$$

If $f:B^n\rightarrow B^n$ is a Boolean function, then it is well known that
there exists a unitary operator $U_f$ which computes all values of $f$ in the
following way:
$$
U_f(\frac{1}{\sqrt{2^n}}\sum_{x=0}^{2^n-1}|x,0\rangle)=
\frac{1}{\sqrt{2^n}}\sum_{x=0}^{2^n-1}U_f(|x,0\rangle)=
\frac{1}{\sqrt{2^n}}\sum_{x=0}^{2^n-1}|x,f(x)\rangle).
$$
On the other hand by well known results (see \cite{bry1}) there
exist unitary matrices $U_2\in U(2)$ and $U_4\in U(4)$ such that
each unitary matrix from $U(2^n)$ can be represented as a product
of matrices $\mathbb{I}\otimes...\otimes
U_2\otimes...\otimes\mathbb{I}$ and $\mathbb{I}\otimes...\otimes
U_4\otimes...\otimes\mathbb{I}$. Such $U_2§, §U_4$ are called
universal gates. Our aim is to represent the universal gates as
S-matrices.

\textbf{Example 1.} The so called Hadamard matrix $ H=\frac{1}{\sqrt{2}}\left
(
\begin{array}{cccc}
1&1 \\
1& -1
\end{array}
\right )
 $
can be represented as $\tau\left (
\begin{array}{cccc}
\sqrt{2}&1 \\
1 & \sqrt{2}
\end{array} \right ).
 $
Similarly the matrix
$
\left (
\begin{array}{cccc}
0&1 \\
1& 0
\end{array}
\right ) $ has a representation
$\tau\left (
\begin{array}{cccc}
1&0 \\
0&1
\end{array} \right ).
 $

\textbf{Example 2.} The complex numbers $a=\sqrt{n^2+1},b=n$ define
$$
\tau\left (
\begin{array}{cccc}
\sqrt{n^2+1}&n \\
n & \sqrt{n^2+1}
\end{array}
\right )= \frac{1}{\sqrt{n^2+1}}\left (
\begin{array}{cccc}
n &1 \\
1 &-n
\end{array}
\right ).
 $$

As $n\rightarrow\infty$ this matrix approximates the universal
gate $ NOT=\left (
\begin{array}{cccc}
1&0 \\
0 & -1
\end{array}
\right ).
 $
Similarly we can approximate more general gate $\left (
\begin{array}{cccc}
1&0 \\
0 & e^{i\phi}
\end{array}
\right ), $ if we take $a=\sqrt{n^2+1},$ $b=-ne^{i\phi/2}.$
Indeed, we have
$$T=\frac{1}{a}=\frac{1}{\sqrt{n^2+1}}e^{i\phi/2},\;\;
\frac{b}{a}=\frac{n}{\sqrt{n^2+1}}e^{i\phi},\;\;
-\frac{\overline{b}}{a}=\frac{n}{\sqrt{n^2+1}}$$ and
 $$
\left (
\begin{array}{cccc}
\frac{1}{\sqrt{n^2+1}}&\frac{1}{\sqrt{n^2+1}}e^{i\phi/2} \\
\frac{1}{\sqrt{n^2+1}}e^{i\phi/2} &\frac{n}{\sqrt{n^2+1}}e^{i\phi}
\end{array}
\right ) \rightarrow \left (
\begin{array}{cccc}
1 &0 \\
0 &e^{i\phi}
\end{array}
\right ),\;\;as\;\; n\rightarrow\infty.
 $$
 \qed

It is known that the Hadamard matrix $H$ and controlled phase gate $ \left (
\begin{array}{cccc}
1&0&0&0 \\
0&1&0&0\\
0&0&1&0\\
0&0&0&e^{i\phi}
\end{array}
\right )
 $
can be used as a universal set of gates, as well as $H$ and
controlled NOT $cNOT =\left (
\begin{array}{cccc}
1&0&0&0 \\
0&1&0&0\\
0&0&0&1\\
0&0&1&0
\end{array}
\right )
 $.

{\bf Remark 2.} There exists a mapping $SU(1,1)\rightarrow SU(2)$
$$
\left (
\begin{array}{cccc}
a&\overline{b} \\
\overline{b}&a\\
\end{array}
\right )\mapsto \left (
\begin{array}{cccc}
\frac{1}{a}&\frac{b}{\overline{a}} \\
-\frac{\overline{b}}{a}&\frac{1}{\overline{a}}\\
\end{array}
\right ),
 $$
 which is used in some problems of differential geometry
 \cite{novikov} but neither $S$ nor $\widetilde{S}$ are generated by this mapping.

 Now we show why the universal gates can be constructed
 via a potential  $Q(x)$ and
particle's momentum $k.$

 Scattering data of the equation (\ref{sch}) are defined as a set
 $\mathcal{S}^{+}=\{R(k)=\frac{b(k)}{a(k)},(k_j,b_j)_{j=1}^N
 \},$ where $R(k)$ is a meromorphic function in the upper half plane, $k_j=i\eta_j,j=1,...,N$
is the finite set of poles of $R(k)$ and
$b_j=\frac{\varphi(x,i\eta_j)}{\psi(x,i\eta_j)}$.
 From these
 scattering data $\mathcal{S}^{+}$ we can recover the scattering matrix,
 i.~e. its unknown element $T(k)=\frac{1}{a(k)}$ in the following way
 $$
 -\ln T(k)=\ln a(k)=\sum_{j=1}^N\ln
 \frac{k-i\eta_j}{k+i\eta_j}-\frac{1}{2\pi
 i}\int_{\mathbb{R}}\frac{\ln(1-|R(\zeta)|^2)}{\zeta-k-i0}d\zeta
 $$
 or
 $$
 T(k)=\prod_{j=1}^N\frac{k+i\eta_j}{k-i\eta_j}e^{\frac{1}{2\pi i}\int_{\mathbb{R}}
 \frac{\ln(1-|R(\zeta)|^2)}{\zeta-k-i0}d\zeta}
 $$
Since by the Plemelj formula $\frac{1}{\zeta-k-i0}={\rm
v.p.}\frac{1}{\zeta-k}+i\pi\delta(\zeta-k)$ (\cite{fadeev}. 50p),
then
$$
 -\ln T(k)=-\frac{1}{2}\ln (1-|R(k)|^2)+\sum_{j=1}^N\ln
 \frac{k-i\eta_j}{k+i\eta_j}-\frac{1}{2\pi
 i}{\rm v.p.}\int_{\mathbb{R}}\frac{\ln (1-|R(\zeta)|^2)}{\zeta-k}d\zeta
 $$
and finally we have
$$
 T(k)=\sqrt{1-|R(k)|^2}\prod_{j=1}^N\frac{k+i\eta_j}{k-i\eta_j}\exp({\frac{1}{2\pi i}
 {\rm v.p.}\int_{\mathbb{R}}
 \frac{\ln(1-|R(\zeta)|^2)}{\zeta-k}d\zeta})
 $$
 Evidently
$|T(k)|=\sqrt{1-R(k)^2}$, since $|k+i\eta|=|k-i\eta|$ and
$\frac{1}{2\pi i}
 {\rm v.p.}\int_{\mathbb{R}}
 \frac{\ln(1-|R(\zeta)|^2)}{\zeta-k}d\zeta$ belongs to $[0,1]$.
Thus we obtain that the phase of $T(k)$ is defined by the Hilbert transform
of the function $\ln(1-R(\zeta)|^2).$ Suppose now the sets of numbers
$\{k_1,...,k_N\},\{\alpha_1,...,\alpha_N\}$ and $\{\beta_1,...,\beta_N\}$ are
given and for simplicity assume $k_j>0$. One can find a continuous function
$f(k),k>0$ such that $f(k_j)=\alpha_j$ and
$\gamma_j\doteq\frac{1}{\pi}\int^{\infty}_0\frac{f(\zeta)}{k_j-\zeta}d\zeta>\beta_j$.
Also we can choose a continuous $g(k)$, $k\le0$ such that
$\frac{1}{\pi}\int_{-\infty}^0\frac{g(\zeta)}{k_j-\zeta}d\zeta=\beta_j-\gamma_j$.
Thus the function
$$
F(k)=
\begin{cases}
f(k),k>0\\
g(k),k\le0
\end{cases}
$$
acquires the property
$$
F(k_j)=\alpha_j,\;\;\;\;\frac{1}{\pi}
\int_{\mathbb{R}}\frac{F(\zeta)}{k_j-\zeta}d\zeta=\beta_j.
$$
Using those relations for a given sequence $k_j,t_j,r_j,j=1,..,N$, such that
$|t_j|^2=|r_j|^2$ we can construct a pair of functions $(T(k),R(k))$ such
that $T(k_j)=t_j,R(k_j)=r_j,$ since for $\alpha_j=t_j,\beta_j=\ln(1-|r_j|^2)$
we find $F(k)$ and $T(k)=\frac{1}{\pi}{\rm v.p.}
\int_{\mathbb{R}}\frac{F(\zeta)}{k_j-\zeta}d\zeta=\beta_j $ and further
define $R(k)$ by the equation $|R(k)|^2=1-e^{-F(k)}$.

The potential $Q(x)$ can be recovered in the following way (see
\cite{newell}). If
$$
C(z)=-i\sum_{j=1}^N\gamma_je^{-\eta_j}+\frac{1}{2\pi}\int_\mathbb{R}
R(\xi)e^{i\xi x}d\xi,
$$
where $\gamma_j=\frac{b_j}{a'(i\eta_j)}$, then $Q$ is defined by the solution
of the Gelfand-Levitan integral equation
$$
K(x,y)+C(x+y)+\int_x^\infty K(x,s)C(s+y)ds=0,\;\;y>x
$$
as $Q(x)=2\frac{d}{dx}K(x,x)$.

Therefore we have proved

{\bf Proposition 1.} Let $k_1,k_2,...,k_m$ be a sequence of pairwise distinct
numbers and $S_1,S_2,...,S_m$ be matrices from the set $\tau(SU(1,1))$. Then
there exists a potential $Q(x)$ such that the corresponding scattering matrix
satisfies the conditions $S(k_1)=S_1,S(k_2)=S_2,...,S(k_m)=S_m$.

\

Now consider a particle with spin moving on the line in presence of an
electromagnetic field $(E(x),B(x)), x\in \mathbb{R}$, where the electric
field $E(x)$ is directed along the line and magnetic field $B(x)$ is directed
across the line.

Let $A(x)$ be the potential of the magnetic field $B(x)$ and $Q(x)$ be the
potential of the electric fields, i.~e. $A^{\prime}(x)=B(x)$,
$Q^{\prime}(x)=-E(x).$

The Schr\"odinger operator (as Hamiltonian) of such system has the form
\begin{equation}\label{magn}
\mathcal{H}=-(\frac{d^2}{dx^2}-A^2(x)+Q(x))\mathbb{I}-B(x)\sigma_3,
\end{equation}
where $\mathbb{I}=\left(
\begin{array}{cccc}
1&0\\
0&1\\
\end{array}
\right )$, and $\sigma_3=\left(
\begin{array}{cccc}
1&0 \\
0&-1\\
\end{array}
\right )$ is third Pauli matrix.

Here for
$
\psi(x)= \left(
\begin{array}{cccc}
\psi_{1}(x)\\
\psi_2(x)\\
\end{array}
\right )\in L^2(\mathbb{R})\otimes \mathbb{C}$
we have
$$
\mathcal{H}\psi=-\left (
\begin{array}{cccc}
\psi''_1(x)+(Q(x)+A^{\prime}(x)-A(x)^2)\psi_1(x)\\
\psi''_2(x)+(Q(x)-A^{\prime}(x)-A(x)^2)\psi_2(x)\\
\end{array}
\right ).
$$

{\bf Remark 3}. Maybe it is more plausible to consider the behavior particle
in the plane. In this case  the potential of magnetic field is of the form
$a(x,y)=(a_1(x,y),a_2(x,y))$ and in the gauge $a_1=0$ we can take
$a'_2(x)=B(x)$ . Then Hamiltonian can be written as
$$
\mathcal{H}=(-\frac{\partial^2}{\partial x^2}-\frac{\partial^2}{\partial
y^2}+ia_2(x)\frac{\partial}{\partial
y}-Q(x)+a_2^2(x))\mathbb{I}-B(x)\sigma_3.
$$
If we take $A(x)=a_2(x)$ and consider  the wave functions
independent on $y$-variable then we get the Hamiltonian
(\ref{magn}).

If in advance we take two arbitrary potentials $U(x)$ and $V(x)$ then $A(x)$
and $Q(x)$ defined by relation
$$
A^{\prime}(x)=\frac{1}{2}(U(x)-V(x)), \text{  }
Q(x)=A^2(x)+\frac{1}{2}(U(x)+V(x))
$$
satisfy the equations
$$
Q(x)+A^{\prime}(x)-A^2(x)=U(x), \text{  }
Q(x)-A^{\prime}(x)-A^2(x)=V(x).
$$
Hence we have
$$
-\mathcal{H}\psi(x)=\frac{d^2}{dx^2}\psi(x)+\left (
\begin{array}{cccc}
U(x)&0\\
0&V(x)\\
\end{array}
\right )\psi(x)\equiv-\left (
\begin{array}{cccc}
H_U\psi_1(x)\\
H_V\psi_2(x)\\
\end{array}
\right ).$$

By definition of the scattering operator it follows that the scattering
operator for the Hamiltonian $\mathcal{H}=\left (
\begin{array}{cccc}
H_U&0\\
0&H_V\\
\end{array}
\right )$ has the form  $S=\left (
\begin{array}{cccc}
S_U&0\\
0&S_V\\
\end{array}
\right ),$ where $S_U$ and $S_V$ are scattering operators for $H_U$ and $H_V$
respectively.

Therefore on the $k^2$-energy level the scattering matrix of the system
(\ref{magn}), which is a $4\times 4$-matrix, can be represented as
 \begin{equation}\label{mtr}
 S_H(k)=\left (
\begin{array}{cccc}
S_U(k)&0\\
0&S_V(k)\\
\end{array}
\right ),
\end{equation}
where $S_U(k)$ and $S_V(k)$ are
scattering matrices corresponding to the potentials of $U$ and $V$
respectively. Evidently they are $2\times 2$-matrices from
$\tau(SU(1,1)).$

Therefore it is proved

{\bf Proposition 2}. Let  $U(x)$ and $V(x)$ be two arbitrary potentials. Then
there exists a electro-magnetic potential
 $(A(x),Q(x))$ such that the Hamiltonian of (\ref{magn}) takes the
 form $\mathcal{H}=\left (
\begin{array}{cccc}
H_U&0\\
0&H_V\\
\end{array}
\right )$, where $H_U=-\frac{d^2}{dx^2}-U(x)$ and
$H_V=-\frac{d^2}{dx^2}-V(x)$. Moreover the scattering matrix for the
stationary  Schr\"odinger equation $\mathcal{H}\psi=k^2\psi$ can be
represented as a matrix (\ref{mtr}).

As a corollary we get that $4\times 4$-matrix from the example 2
$$
\left (
\begin{array}{cccc}
1&0&0&0 \\
0&1&0&0\\
0&0&1&0\\
0&0&0&e^{i\phi}
\end{array}
\right )
 \text{and}
\left (
\begin{array}{cccc}
1&0&0&0 \\
0&1&0&0\\
0&0&0&1\\
0&0&1&0
\end{array}
\right ) $$ can be represented as  the scattering matrix of the system
governed by (\ref{magn}). Moreover we can take the potential $U(x)=0$ in both
cases and $V(x)$ as indicated in example 1 and 2.

\section{Two-level system in electric field}

Let us consider  behavior of a two-level system in an electric
 field. It can be described by the equation
\begin{equation}\label{s1}
i\frac{d}{dt} \left (\begin{array}{cccc}
A(t)\\
B(t)
\end{array}
\right ) = \left (\begin{array}{cccc}
\zeta&E(t)\\
\overline{E(t)}&-\zeta\\
\end{array}
\right )\left (\begin{array}{cccc}
A(t)\\
B(t)
\end{array}
\right )
\end{equation}
where  $A$ and $B$ denote the amplitudes of first and second level
respectively, $E(t)$ is the complex envelope of the electric field and
$\zeta$ is the difference between support frequency and proper frequency. The
free Hamiltonian here is $H_0=\zeta\sigma_3$ and free dynamics is given by
the unitary matrix $e^{-i\zeta t\sigma_3}=\left(\begin{array}{cccc}
e^{-i\zeta t} & 0 \\
0 &e^{i\zeta t}
\end{array}
\right)$.

Let $\Psi(t)$ and $\Phi(t)$ be matrix solution of (\ref{s1}) with
condition
\begin{equation}\label{c1}
\Psi(t)\sim\left (
\begin{array}{cccc}
e^{-i\zeta t} & 0 \\
0 &e^{i\zeta t}
\end{array}
\right ), t \rightarrow \infty \text{ and } \Phi(t)\sim\left (
\begin{array}{cccc}
e^{-i\zeta t} & 0 \\
0 &e^{i\zeta t}
\end{array}
\right ), t \rightarrow -\infty
\end{equation}
respectively. They are unitary unimodular matrices, since $\left(
\begin{array}{cccc}
{\zeta } & E(t) \\
\overline{E(t)} &{-\zeta}
\end{array}
\right )$ is a Hermitian traceless matrix. Thus the so called monodromy
matrix $M(\zeta)=\Psi^{-1}(t)\Phi(t)$ belongs to $SU(2)$ and can be
represented as $M(\zeta)=\left (
\begin{array}{cccc}
a(\zeta) & -\overline{b}(\zeta) \\
b(\zeta) &\overline{a}(\zeta)
\end{array}
\right )$ with $|a(\zeta)|^2+|b(\zeta)|^2=1.$

Let $U(t,\tau)$ be the fundamental matrix of (\ref{s1}), i.~e.
$$
U(\tau,\tau)=I,\;\; i\frac{\partial}{\partial t }U(t,\tau)=\left
(\begin{array}{cccc}
\zeta&E(t)\\
\overline{E(t)}&-\zeta\\
\end{array}
\right )U(t,\tau),\;\;t>\tau.
$$

Evidently there exist matrices $W_\varphi$ and $W_\psi$ such that
$\Psi(t)=U(t,0)W_{\Psi},$ $\Phi(t)=U(t,0)W_{\Phi}.$ The relation (\ref{c1})
can be rewritten as
$$
\lim_{t\rightarrow \infty}\Psi(t)^{\star}e^{-\zeta t
\sigma_3}=\mathbb{I} \;\;\text{and}\;\; \lim_{t\rightarrow
-\infty}\Phi(t)^{\star}e^{-\zeta t \sigma_3}=\mathbb{I}.
$$
On the other hand by definition of wave operators we have
\cite{reed}
$$
W_{\pm}=\lim_{t\rightarrow \mp \infty}U(t,0)^* e^{-\zeta
t\sigma_3}.
$$
Hence
$$
W_+=\lim_{t\rightarrow -\infty}U(t,0)^* e^{-\zeta t\sigma_3}
=\lim_{t\rightarrow -\infty}W_\Phi(U(t,0)W_\Phi)^* e^{-\zeta
t\sigma_3}
$$
$$=\lim_{t\rightarrow -\infty}W_\Phi\Phi(t)^*
e^{-\zeta t\sigma_3}=W_\Phi.
$$
Similarly $W_-=W_\Psi$. Thus we have
$$
S(\zeta)=W^{-1}_-W_+=W_\Psi^{-1}W_\Phi
$$
$$
=\Psi(t)^{-1}U(t,0)U(t,0)^{-1}\Phi(t)=
\Psi(t)^{-1}\Phi(t)=M(\zeta)
$$
i.e.
$$
S(\zeta)=M(\zeta)=\left (
\begin{array}{cccc}
a(\zeta) & -\overline{b}(\zeta) \\
b(\zeta) &\overline{a}(\zeta)
\end{array}
\right ),\;\;|a(\zeta)|^2+|b(\zeta)|^2=1.
$$
The function $a(\zeta)$ is the boundary value of a holomorphic function on
the upper half plane with zeros $\zeta_{j},$
 $Im\zeta_j>0,$ $j=1,...,N$. For $\zeta=\zeta_j$ there exists a
 solution $\chi(t)$ of (\ref{s1}) with condition $\chi(t)\sim\left (
\begin{array}{cccc}
0\\
e^{-k_jt}
\end{array}
\right ),$ $t\rightarrow -\infty,$ $\chi(t)\sim\left (
\begin{array}{cccc}
e^{ik_jt}\\
0
\end{array}
\right )d_j,$ $t\rightarrow \infty.$ The scattering data for the system
(\ref{s1}) is the set $\{ b(\zeta),\zeta_j,d_j,j=1,...,N \}.$ The function
$a(\zeta)$ is defined as
$$
a(\zeta)=e^{-\frac{1}{2\pi
i}\int_{\mathbb{R}}\frac{\ln(1-|b(\zeta^{\prime})|^2)}
{\zeta-\zeta^{\prime}-i0}d\zeta^{\prime}}
\prod_{l=1}^{N}\frac{\zeta-\zeta_l}{\zeta-\overline{\zeta}_l}.
$$
The field $E(t)$ can be restored by scattering data as follows (see
\cite{fadeevl}); $E(t)=-2iK(t,t),$ where $K(t)=\left(
\begin{array}{cccc}
K_{11}(t) & K_{12}(t) \\
K_{21}(t) &K_{22}(t)
\end{array}
\right)$ is a solution of the Gelfand-Levitan integral matrix equation
$$
K(t,s)+\widehat{F}(t,s)+\int_{t}^{\infty}K(t,s)\widehat{F}(z+s)dz,\;\;t<s
$$
with
$$\widehat{F}(t)= \left (
\begin{array}{cccc}
0&F(t) \\
\overline{F}(t) & 0
\end{array}
\right ), F(t)=\frac{1}{2\pi}\int_{\mathbb{R}}r(k)e^{ikt}dk
+\sum_{j=1}^{N}m_{j}e^{ik_jt},
$$
$$r(\zeta)=\frac{b(\zeta)}{a(\zeta)},\;\;
m_{j}=\frac{d_j}{a^{\prime}(\zeta_j)}.
$$

As in previous case we define the function $b(\zeta)$ to obtain a desired
scattering matrix $S(\zeta).$ Hence we have

{\bf Proposition 3.} For the finite set of numbers
$\{\zeta_1,\zeta_2,...,\zeta_m\}$ and matrices $S_1,S_2,...,S_m$ from the set
$SU(2)$ there exists a potential $Q(x)$ such that the corresponding
scattering matrix satisfies the conditions
$S(\zeta_1)=S_1,S(\zeta_2)=S_2,...,S(\zeta_m)=S_m$.

In particular cases we can construct the matrices of example 1 up to a
constant:
 $iH=\frac{i}{\sqrt 2}\left(
\begin{array}{cccc}
1 & 1 \\
1 & -1
\end{array}
\right ),\;{\rm and}\;e^{-i\varphi/2}\left (
\begin{array}{cccc}
1 & 0 \\
0 & e^{i\varphi}
\end{array}
\right ).$

Now we suppose that electric field acts on the pair of 2-level particles
performing dipole-dipole interaction between dipolar particle states. We
assume that the dipoles are oriented along electric fields and therefore the
hamiltonian of the two-particle system is presented as
$\mathcal{H}_{AB}(E(t),\Phi_{AB})$ \cite{Lee}, where
$$
\mathcal{H}_{AB}(x,y)=(\mathcal{H}_A+x\hat d_A)\otimes I+I\otimes(\mathcal{H}_B+
x\hat d_B)+y\hat d_A\otimes\hat d_B
$$
and
$$
\mathcal{H}_C=\left (
\begin{array}{cc}
W_{+}^C&0 \\
0 & W^C_{-}
\end{array}
\right ) \;\;
 \hat d_C=\left (
\begin{array}{cccc}
0&d_C \\
\bar d_C & 0
\end{array}
\right )\;\; C=A,B
$$
In the case of $\Phi_{AB}=y$ the Hamiltonian can be rewritten as
$4\times 4$ matrix
$$
\mathcal{H}_{AB}(E(t),y)=\left (
\begin{array}{cccc}
W_{+}^A+W_{+}^B&d_AE(t)& d_AE(t) &yd_Ad_B \\
\bar d_AE(t) & W^A_{+}+W_{-}^B & yd_A\bar d_B&d_BE(t)\\
\bar d_AE(t)&y\bar d_Ad_B&W_{-}^A+W_+^B&d_BE(t)\\
y\bar d_A\bar d_B&\bar d_B E(t)&\bar d_BE(t)&W_{-}^A+W_-^B
\end{array}
\right ).
$$
{\bf Proposition 4}. Let $d_A,d_B,W_{+}^A,W_{+}^B,W_{-}^A,W_{-}^B $ be
arbitrary complex numbers. Then there exists a continuous electric pulse
$E(t)$ and interacting potential $\Phi_{AB}(t)$ vanishing at infinity and
such that the corresponding scattering matrix does not belong in
$SU(2)\otimes SU(2)$ i.~e. it is entangled operator.

{\it Proof}. It is easy to see that the hamiltonian $\hat
d_A\otimes\hat d_B$ is not represented as a matrix $M\otimes
I+I\otimes N$ for some $M,N\in su(2)$. Hence the matrix
\begin{equation}\label{sca}
F(T)=\exp\{iT\mathcal{H}_{AB}(0,0)\}\exp\{-2iT\mathcal{H}_{AB}(x,y)\}
\exp\{iT\mathcal{H}_{AB}(0,0)\}
\end{equation}
for some $T$ and $y\neq 0$ can not be represented as an element of
$SU(2)\otimes SU(2)$, as
$$
F'(0)=i\mathcal{H}_{AB}(0,0)-2i\mathcal{H}_{AB}(x,y)
+i\mathcal{H}_{AB}(0,0)
$$
also cannot be represented as a matrix $M\otimes I+I\otimes N$ for some
$M,N\in su(2)$. It is easy to see that the S-matrix of the system
$$
i\frac{d}{dt}Y(t)=\mathcal{H}_{AB}(E(t),\Phi_{AB}(t))Y(t)
$$
for the electric pulse $E(t)=x1_{[-T,T]}(t)$ and
$\Phi_{AB}(t)=y1_{[-T,T]}(t)$ coincides with matrix of (\ref{sca}).

Since $SU(2)\otimes SU(2)$ is a closed subset of $U(4)$, for sufficiently
small nonzero $x$ the matrix (\ref{sca}) also is an entangled one, i.~e. does
not belong to $SU(2)\otimes SU(2)$, thus each continuous pulse
$E_\varepsilon(t)$ $\Phi_{AB,\varepsilon}(t)$ close enough to
$E(t),\Phi_{AB}(t)$ also generates an entangled operator. \qed

This proposition and results of \cite{bry1} allow us to tune the electric
pulse to construct a system of universal gates.

\section{Monodromy of Fuchsian systems and the S-matrix}

Let $\{ s_1,...,s_n\}$ be complex numbers on $\mathbb{C}$ and $s_i\neq s_j,$
$i\neq j.$ Consider the Fuchsian system
\begin{equation}\label{f}
df=\omega f,
\end{equation}
where $\omega$ is a meromorphic 1-form on $\mathbb{C}$
$$
\omega=\left(\frac{A_1}{z-s_1}+\frac{A_2}{z-s_2}+...+\frac{A_n}{z-s_n}\right)dz,
$$
and $A_j,$ $j=1,...,n$ are constant $2\times 2$-matrices. Let $(f_1,f_2)$ be
solutions  of (\ref{f}) in the neighborhood of $z_0\in \mathbb{C}\setminus \{
s_1,...,s_n\}$ and $\gamma_1,...,\gamma_n$ be the generators of
$\pi_1(\mathbb{C}\setminus \{s_1,...,s_n\},z_0).$ After the extension of
$(f_1,f_2)$ along the $\gamma_j$ we obtain other solutions $(f_1^j,f_2^j)$ of
(\ref{f}) and $(f_1,f_2)=M_j(f_1^j,f_2^j),$ where $M_j,$ $j=1,...,n$ are
monodromy matrices. Conversely, for fixed data $(s_1,...,s_n, M_1,...,M_n)$
there exists a Fuchsian system (\ref{f}) with given singular points $s_j,$
$j=1,...,n$ and monodromy matrices  $M_j,$ $j=1,...,n.$ We take monodromy
matrices from $SU(2)$ and consider them as scattering matrices. The singular
points $s_1,...,s_n$ are considered as sources of energy  and $\omega$ is the
gauge potential induced from the given data $(s_1,...,s_n, M_1,...,M_n)$.

Now we intend to show how the S-matrix of the system may be expressed as a
monodromy matrix of the corresponding Fushian system. First we recall that
two differential equations $i\frac{d}{dt}\psi_1(t)=\mathcal{H}_1(t)\psi_1(t)$
and $i\frac{d}{dt}\psi_2(t)=\mathcal{H}_2(t)\psi_2(t)$ are called gauge
equivalent if there exists a differentiable 1-parameter family of invertible
matrices such that
$\mathcal{H}_2(t)=iU'(t)U(t)^{-1}+U(t)\mathcal{H}_1(t)U(t)^{-1}$. It is easy
to see that if $\psi_1(t)$ is a solution of the first equation then
$\psi_2(t)=U(t)\psi_1(t)$ is the solution of the second equation. By this
definition the Schr\"odinger  equation of type
$i\frac{d}{dt}\psi_0(t)=(\mathcal{H}_0+V(t))\psi_0(t)$ is gauge equivalent to
the so-called interacting representation
$i\frac{d}{dt}\psi_1(t)=V_1(t)\psi_1(t)$ via 1-parameter family
$U(t)=e^{it\mathcal{H}_0}$, where
$V_1(t)=e^{it\mathcal{H}_0}V(t)e^{-it\mathcal{H}_0}$.

Now we consider a field of type $E(t)e^{-i\omega t}$. Then the system
(\ref{s1}) recasts
\begin{equation}\label{s10}
i\frac{d}{dt} \left (\begin{array}{cccc}
A(t)\\
B(t)
\end{array}
\right ) =\zeta\sigma_3+\left (\begin{array}{cccc}
0&E(t)e^{-i\omega t}\\
{E(t)e^{i\omega t}}&0\\
\end{array}
\right )\left (\begin{array}{cccc}
A(t)\\
B(t)
\end{array}
\right ).
\end{equation}
Since
$$
\left (\begin{array}{cccc}
e^{i\zeta t}&0\\
0&e^{-i\zeta t}\\
\end{array}
\right )\left (\begin{array}{cccc}
0&e^{-i\omega t}\\
e^{i\omega t}&0
\end{array}
\right )\left (\begin{array}{cccc}
e^{-i\zeta t}&0\\
0&e^{i\zeta t}\\
\end{array}
\right )=\left (\begin{array}{cccc}
0&e^{-i(\omega-2\zeta) t}\\
e^{i(\omega t-2\zeta)}&0
\end{array}
\right ),
$$
the interacting representation of the system ({\ref{s10}) takes the form
\begin{equation}\label{s11}
i\frac{d}{dt}\Phi(t)=\left (\begin{array}{cccc}
0&E(t)e^{-i(\omega-2\zeta)t}\\
E(t)e^{i(\omega-2\zeta)t}&0
\end{array}
\right )\Phi(t)
\end{equation}
Let $\Phi(t,\tau)$ denote the fundamental matrix of ({\ref{s11}). Then the
S-matrix of the system ({\ref{s10}) can be represented as
\begin{equation}\label{sca1}
S(\zeta)=\lim_{\begin{array}{cc}t\to+\infty,\\
\tau\to-\infty\end{array}}\Phi(t,\tau).
\end{equation}
In the particular case of analyticity of the function $E(t)$ the S-matrix
coincides with the monodromy matrix of the corresponding Fushian system

{\bf Example 3.} Consider the case
$\omega=2\zeta,\;\;E(t,a,b)=\frac{2ab}{t^2+a^2}\equiv\frac{ib}{t+ia}-\frac{ib}{t-ia},\;\;a,b\in
\mathbb{R} $. By the equality $\sigma_1=H\sigma_3H^{-1}$ the system
(\ref{s11}) is gauge equivalent to
$$
i\frac{d}{dt}\Psi(t)=\frac{2ab}{t^2+a^2}\sigma_3\Psi(t).
$$
The solution of the last equation is
$$
\Psi(t)=\left(\begin{array}{cccc}\exp(-ib\int\frac{2adt}{t^2+a^2})&0\\
0&\exp(ib\int\frac{2adt}{t^2+a^2})\end{array}\right)
$$
and thanks to the formula (\ref{sca1}) the  S-matrix is $e^{-2\pi
b\int\frac{2adt}{t^2+a^2}i\sigma_3}=e^{-bi\sigma_3}$. Since
$\Phi(z)=H\Psi(z)$, one has
$$S(\zeta)=M(\gamma)=H\left
(\begin{array}{cccc}
e^{2\pi bit}&0\\
0&e^{-2\pi bit}
\end{array}
\right )H^{-1}=e^{-2i\pi b\sigma_1}.
$$
Now extend the range of the variable $t$ to the complex plane $\mathbb{C}$
and compute the monodromy matrix of the so obtained Fuchsian system. The
system (\ref{s11}) looks as follows
$$
i\frac{d}{dw}\Phi(w)=\frac{2ab\sigma_1}{w^2+a^2}\Phi(w),
$$
where $\sigma_1=\left(\begin{array}{cccc}
0&1\\
1&0
\end{array}\right)$.
By the equation $\sigma_1=H\sigma_3H^{-1}$ this system is gauge
equivalent to
$$
i\frac{d}{dw}\Psi(w)=\frac{2ab\sigma_3}{w^2+a^2}\Psi(w).
$$
 The line $Im w=0$ on the Riemann sphere
$P_1(\mathbb{C})\supset \mathbb{C}$ becomes a loop. In order to express this
loop explicitly we make change of coordinate $z=\frac{1}{w-i},$ which is a
holomorphic map $P_1(\mathbb{C})\to P_1(\mathbb{C})$ with inverse
$w=\frac{1}{z}+i$. This mapping carries the 1-form $\frac{2ab}{w^2+a^2}dw$ to
the 1-form $-\frac{2ab}{a^2z^2+(iz+1)^2}dz$. In view of
$$-\frac{2ab}{a^2z^2+(iz+1)^2}=\frac{ib}{z-\frac{i}{a+1}}-\frac{ib}{z+\frac{i}{a-1}}$$
we get the Fushian system
\begin{equation}\label{fush}
\frac{d}{dz}\Psi(z)=\left(\frac{b\sigma_3}{z-\frac{i}{a+1}}-\frac{b\sigma_3}{z+\frac{i}{a-1}}
\right)\Psi(z),
\end{equation}
which has the solution
$$
\Psi(z)=\left(\begin{array}{cccc}
\exp\left(b\int\frac{dz}{z-\frac{i}{a+1}}-b\int
\frac{dz}{z+\frac{i}{a-1}}\right)&0\\
0&\exp\left(-b\int\frac{dz}{z-\frac{i}{a+1}}
+b\int\frac{dz}{z+\frac{i}{a-1}}\right) \end{array}\right).
$$
In new coordinates the line $w=t$ becomes the curve
$z=\frac{1}{t-i}\equiv\frac{t}{1+t^2}+i\frac{1}{1+t^2}$, which is circle
$\gamma:x^2+(y-\frac{1}{2})^2=\frac{1}{4}$. The poles of system (\ref{fush})
$\frac{i}{a+1}$ and $\frac{i}{1-a}$ lie inside and outside of the circle
respectively. Hence we can consider the circle $\gamma$ as loop around pole
$\frac{i}{a+1}$ and monodromy matrix of (\ref{fush}) coincides to S-matrix
given by (\ref{sca1}) Since for solution  of (\ref{fush}) is
$\Phi(z)=H\Psi(z)=H^{-1}\Psi(z)$, one has
$$S(\zeta)=M(\gamma)=H\left (\begin{array}{cccc}
e^{2\pi bi}&0\\
0&e^{-2\pi bi}
\end{array}
\right )H^{-1}=e^{-2\pi bi\sigma_1}.$$\qed

If $E(t)=\sum\frac{2a_kb_k}{t^2+a_k^2}$ then $S(\zeta)=\prod M(\gamma_k)$,
where $\gamma_k$ is a loop around $\frac{1}{1+a_k}$ and $M(\gamma_k)$ is the
corresponding monodromy matrix.

{\bf Example 4.} If we consider the field $E(t)=\frac{2t}{t^2+a^2}$, the
integral $\int_{-\infty}^\infty\frac{2tdt}{t^2+a^2}$ exists only in the sense
of principal value. Extending the field in the complex plane, after change of
variable we obtain
\begin{equation}\label{fush1}
\frac{d}{dz}\Phi(z)=
\frac{1}{a}\left(\frac{1}{z}+i\right)\left(\frac{\sigma_1}{z-\frac{i}{a+1}}-
\frac{\sigma_1}{z+\frac{i}{a-1}}
\right)\Phi(z).
\end{equation}
Hence the complementary pole $z=0$ appears, which lies in $\gamma$. We pass
to a gauge equivalent differential equation of type
$$
\frac{d}{dz}\Psi(z)=\left(\frac{b_1}{z}+\frac{b_2}{z-p}+\frac{b_3}{z-q}
\right)\sigma_3\Psi(z),\;\;p=\frac{i}{1+a},\;q=\frac{i}{1-a}
$$
for some $b_1,b_2,b_3\in \mathbb{C}$. The monodromy matrix of such system
around $\gamma$ is the matrix
$$
\left (\begin{array}{cccc}
e^{\int_\gamma\left(\frac{b_1}{z}+\frac{b_2}{z-p}+\frac{b_3}{z-q}\right)dz}&0\\
0&e^{-\int_\gamma\left(\frac{b_1}{z}+\frac{b_2}{z-p}+\frac{b_3}{z-q}\right)dz}
\end{array}
\right )
$$
which contains a Cauchy type integral ${\rm v.p.
}\int_\gamma\frac{b_1}{z}dz$. Since 0 lies on $\gamma$ and $p$ lies inside
$\gamma$, this matrix is equal to
$$\left(\begin{array}{cccc}
\exp(i\pi b_1+2\pi ib_2)&0\\
0&\exp(-i\pi b_1-2\pi ib_2)
\end{array}\right)=\exp(i\pi(b_1+2b_2)\sigma_3).$$ Hence the
system (\ref{fush1}) has the monodromy matrix $\exp(i\pi(b_1+2b_2)\sigma_1)$
similarly to the previous example.

{\bf Remark 4.} In the case of $\omega\neq2\zeta$ the more complicated
irregular singularities emerge. We intend to study them in subsequent works.

\end{document}